\documentclass{article}
\usepackage[dvips]{graphicx}
\begin{document}

\pagestyle{plain} 
\setcounter{page}{1}
\setlength{\textheight}{700pt}
\setlength{\topmargin}{-40pt}
\setlength{\headheight}{0pt}
\setlength{\marginparwidth}{-10pt}
\setlength{\textwidth}{20cm}

\title{Some Considerations on  Six Degrees of Separation from A Theoretical Point of View}
\author{Norihito Toyota \and Hokkaido Information University, Ebetsu, Nisinopporo 59-2, Japan \and email :toyota@do-johodai.ac.jp }
\date{}
\maketitle
\begin{abstract}
In this article we discuss six degrees of separation, which has been proposed by Milgram, from a theoretical point of view.
Simply if one has $k$ friends, the number $N$ of  indirect friends  goes up to  $\sim k^d$ in $d$ degrees of separation. 
So it would easily come up to population of  whole world. 
That, however, is unacceptable. 
Mainly because of nonzero clustering coefficient $C$, $N$ does not become $\sim k^d$.
In this article,  we first discuss relations between six degrees of separation and the clustering coefficient 
in the small world network proposed by Watt and Strogatz\cite{Watt1},\cite{Watt2}. 
Especially,  conditions that $(N)>$ (population of U.S.A  or  of the whole world) arises  in the WS model is explored by theoretical  and numerical points of view. 
Secondly we introduce an index that represents velocity of propagation to the number of friends and obtain an  
analytical formula for it as a function of $C$, $K$, which is an average degree over all nodes, and some parameter $P$ concerned with  network topology. 
Finally the index is calculated numerically to study the relation between $C$, $K$ and $P$  and $N$.      

 \end{abstract}
\begin{flushleft}
\textbf{keywords:}
Six Degrees of separation,  Small world Network, Propagation Coefficient, Watt-Strogatz Model, Clustering Coefficient, 
Average Path Length
\end{flushleft}

\section{Introduction}
In 1967, Milgram made a great impact on  the world by advocating the concept 
 "six degrees of separation" by an social experiment in a celebrated paper \cite{Milg}. 
"Six degrees of separation"  shows that people have a narrow circle of acquaintances. 
A series of social experiments made by him suggest 
that all people in USA are connected through about 6 intermediate acquaintances.   
In this paper we inspect from a rather theoretical point of view that 
this phenomenon, so called "six degrees of separation" is not so surprising and if anything natural one.  

This article is first motivated by a following simple consideration; 
If every person has $K$ acquaintances, so that after $L$ steps of intermediate acquaintances
the person would be able to convey a mail,  in general information, to about $S=K^L$ persons. 
With the proviso, however, that the network of acquaintances has a tree structure without any loop 
( no clustering coefficient), evaluating it in more detail, it is 
\begin{equation}
S= \sum_{i=0}^{L-1} K(K-1)^i =\frac{K(K^{L-1}-1)}{K-1},  
\end{equation}  
where it is assumed that the relation of acquaintances is symmetric.  
Thus information will spread out among exponentially many persons from one person with $L$ steps.     
Though a person that received information, of course, may convey the information 
to only a part of his/her acquaintances, when a network has a tree structure,  
six degrees of separation for any two persons would not be so mysterious.  
 
Real networks, however,  naturally have structures with loops.  
If there is some loops or an effective clustering coefficient in the network of acquaintances, 
this discussion will greatly altered.  
One of aims of this article is to evaluate the effect of clustering coefficient 
on propagation of information.  
How much does the clustering coefficient reduce the population that information is provided ? 
We consider it from three points of view.      
They consist of the following three;

1. To inspect six degrees of separation for Watts-Strogatz type model where analytic expressions 
for the clustering coefficient and the average path length are found. 

2. To inspect six degrees of separation for more general small world networks with uniform clustering coefficient. 

3. To consider empirical networks such as a network in Mixi and a network of actor/actress  besed on data of Bacon game.\\

The plan of this article is as follows. 
In the next section we argue on Watts-Strogatz type small world networks\cite{Watt1,Watt2} 
where theoretical evaluation of the average length and the clustering coefficient has been made. 
We study the propagation of information on the networks by making numerical analyses based on  these formulae,  
To study more general small world networks,   we present a propagation coefficient model in the section 3.
We can not analyze, however, large class of small world networks including scale free networks  
due to technical reasons at present. 
We adopt a homogeneous hypothesis which would be explained in detail in the section 3. 
Various types of homogeneity are postulated in the propagation coefficient model.  
In the section 4, we analyze experimental data, such as a network in Mixi and a network based on data of Bacon game, 
as to estimate clustering coefficient.  
Concluding remarks are given in the last section. 

\section{Watts-Strogatz Model}
In this section we investigate the effect of the clustering coefficient on diffusion of information in 
Watts-Strogatz type small world networks.  
The analytic expressions for the clustering coefficient and the average path length in the networks 
have been found.  
For the original Watts-Strogatz version of the model, the clustering coefficient $C(p)$ has been given 
\cite{Barr1} by  
\begin{equation}
C(p) = \frac{3K-6}{4K-2} (1-p)^3,  
\end{equation}
where $n$ is the network size, that is, the number of nodes on the network, $K$ is 
the average degree and $p$ is the rewiring probability. 
Moreover the average node-node separation $L(p)$ in a modified version of the original Watts-Strogatz model 
has found in the limit of low density of shortcuts\cite{Newm2}; 
\begin{equation}
L(p)=\frac{2n}{K}F(\frac{nKp}{2}),\;\;\;F(x)=\frac{1}{2\sqrt{x^2+2x}}\tanh^{-1}\sqrt{\frac{x}{x+2}}, \;\; \mbox{ for small $p$}. 
\end{equation}
In the modified version of the original small world model in which shortcuts edges are added 
between randomly chosen node pairs, no bonds are removed. 
Here edges are not rewired, rather adding edges thus ensuring that the modified network stays connected.
We attempt analyses of six degrees of separation based on these two formulae.

\subsection{Parameter Regions for Six Degrees of Separation }
In this subsection, we explore possible regions of parameters introduced in the previous section 
for six degrees of separation. 
First we explore  parameters' regions in order that information can spread among 
on average $10^9$ persons by $6\pm 1$ steps. 
Taking $L= 6\pm 1$ and $n=10^9$ in Eq. (3), we explore the possible regions in a $p-K$ space. 
We can numerically find them within meaningful values of the parameters. 
They are shown in Fig.1.  
As far as $p$ is not extremely small that is the small world region, that $K$, 
 the average number of contacts of a person, is several tens 
is sufficient for information to spresd from one person to $10^9$ persons.   

The solution for  $C-K$ space where $p$  is eliminated from  Eq.(1) and Eq.(2)  is described in Fig. 2. 
Though Eq. (3) only holds only at small $p$, then at least we can work out  two equations  simultaneously. 
Since as $p$ becomes large, $C$ becomes small, the validity of the analysis would be lost for small $C$.   
 Fig. 2 asserts that it is sufficient that $K$ is several tens for relatively large values of $C$.    
\begin{center}
\includegraphics[scale=0.8,clip]{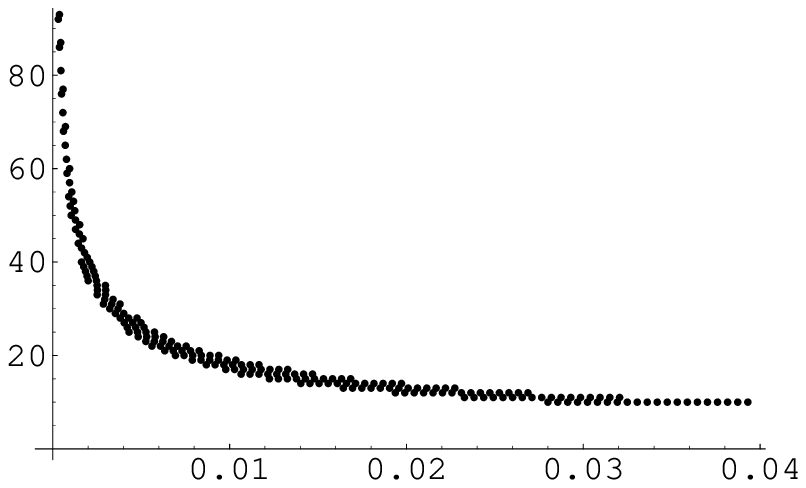} \includegraphics[scale=0.8,clip]{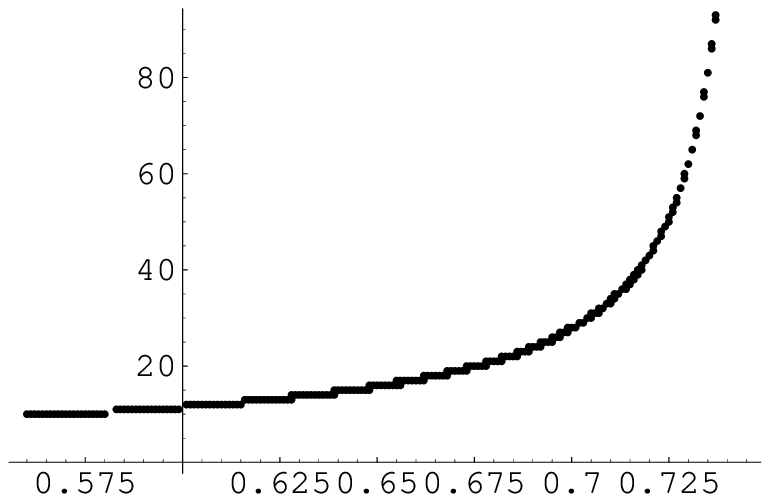}
\\
Fig．1.  p-K plots in SW model.$\;\;\;\;\;\;\;\;\;\;$
Fig．2. C-K plots in SW model. 
\end{center}　


Through these analyses, we conclude that following regions of parameters are roughly needed; 
\begin{eqnarray}
C&=& 0.55\sim 0.75,  \nonumber \\ 
p&=& 0.01 \sim 0.04,  \nonumber \\ 
K&=& 15 \sim 25, \nonumber 
\end{eqnarray}
in order that 
information can spread from one person to $10^9$ persons at about 6 steps on Watts-Strogatz type small world networks.   

\subsection{Population propagated at $L=6$}
In this subsection we consider that to how many persons information can propagate at exact six steps from a person.
We find adequate regions in a parameters' space $K-C$  when   $n$ falls between $10^7$ and $10^9$ at $L=6$. 
For each $K$, typical values of $C$ are listed in Table 1. 

\begin{center}
Table 1.  $C$ and $K$ for $n\sim 10^7\sim 10^{9} $ \\
\vspace{5mm}
\begin{tabular}{|l|c|c|c|c|c|c|c|c|c|c| } \hline 
K   & 9&10&12&15&20 &25 &49 &100& 147& 194    \\ \hline \hline
C   &0.54 & 0.57& 0.61& 0.64&0.68 & 0.69& 0.724& 0.738&0.74&0.744\\ \hline
\end{tabular}
\end{center}

As $K$ becomes large, $n$ obviously becomes large.  
From Table 1, $K$ is not so large even for rather large $C$. 
Thus information can readily spread to about a billion persons at 6 steps, if $C$ took fairly small value. 

\begin{center}
Table 2.  Empirical Networks with large $C>0.5$\\
\begin{tabular}{|c|c|c|c|c|c| } \hline 
Network    & Number of Vertices  & $<K>$  & $C$  &$<L>$ & Index \\ \hline 
Company directors &7673 &14.4 & 0.59 &&\\ \hline
Coauthorships &56627 && 0.726 & 4 &1.2\\
    in the SPIREES e-archive    &&&&& \\ \hline
Collaboration net & 70975 && 0.59 &2.1294 & 9.5\\ 
collected from math.journals  &&&&& \\ \hline
Collaboration net  & 209293 && 0.76 &2.4 & 6\\ 
collected from neurosci.journals   &&&&& \\ \hline
metabolic network  & 315  && 0.59 & &  \\ \hline
World Web                                 & 470000 && 0.69/0.44 &1.5/2.7 &2.65\\ \hline
\end{tabular}
\end{center}

So far we have seen that it has been possible to realize six degrees of separation even rather large $C$. 
There, however, are not so many empirical networks with large $C$. 
Networks with rather large $C$  that have discovered so far are listed in Table 2
\cite{Newm2},\cite{Albe4},\cite{Ferr},\cite{Fell},\cite{Albe4},\cite{Doro2}.   
Here  blanks represent that they are unknown and the index in the rightest column is that for scale free.  
These networks mostly have scale free nature, so that they are not Watts-Strogatz type small world networks. 

\section{Propagation Coefficient Model }
We study Milgram-like-propagation of information in wider class of networks, including those other than small world networks. 
First we focus our attention on any one node, which is the node in  0-generation,  in general networks. 
Next we explore all nodes connected with the first node, which are nodes in the 1-generation.   
Next we explore all nodes connected with the nodes in the 1-generation apart from the nodes of  0-generation, which are nodes in the  2-generation.   
We continue these procedures until all nodes on a network are covered by these procedures. 
These procedures are effective in any complex networks.  
There are only two generations in complete graphs. 
The maximal generation number $N_G$  of a network is larger than the diameter of the network  by 1 
and so the maximal generation number $G$ is just equal to the diameter.  
This picture of networks make the analyses of the propagation of information on the network manageable. 

Now we introduce some geometrical quantities that used in this article together with their notation. \\

$G_i$ means the i-th generation and $n_i$ is the numbers of generation $G_i$. \\

$N$ is the total number of nodes of a network or the size of the network. \\

$ k^{(j)}_{i,i+1}$ is  the number of edges  from node $j$ in $G_i$  to nodes in $ G_{i+1} $. \\

$ k_{i,i} $ is the number of edges  connected between the same generation  $G_i$.  \\

$C_i$ is a contribution to the clustering coefficient produced by edges in $G_i$. \\

$K^{(j)} $is the degree of the node $j$. \\
We define the propagation coefficient from $G_i$ to $G_{i+1}$ by $ \overline{k}_{i,i+1} $ as  an average of $ k_{i,i+1}$.

 We make the assumptions for simplicity of analyses. \\
1. The  size of the network is infinite.\\
2. A parameter $q_j$ is the possibility  that a node $j$ has two parents. \\
3. There is no backflow in the propagation of information. \\
4. The homogeneous hypothesis ;
\begin{eqnarray}
&&C_i = \overline{C}=const.\\
 &&q_j=q=const. \;\;\; but\;\;\;  q=0 \;\; at \;\; G_0 \;\;generation.\\
&&K^{(j)} =K = const \;\;\;(nearly \;\;equal \;\;degree).
\end{eqnarray}

Under these assumptions, the following relations hold; 
\begin{eqnarray}
\overline{k}_{i,i+1} &=& \frac{\sum_{j\in G_i}^{n_i} k^{(j)}_{i,i+1}}{n_i},  \\
K  &=& 1+\frac{2k_{i,i}}{n_i} +  \overline{k}_{i,i+1}, \\
n_{i+1} &=&   \overline{k}_{i,i+1}  n_i (1-q),  \\
N&=&\sum_{i}^d n_i.
\end{eqnarray}

In order to investigate the effect of the clustering coefficient upon the propagation of information on a network, 
we consider when does the clustering coefficient increase in this picture of networks. 
Notice that there are two cases that make a contribution to the clustering coefficient in this picture of networks. 
One is the case that ages are linked together with nodes in the same generation. 
The other is the case that one node has two parent nodes that are linked each other. 
They are shown in the Fig.3. 
  
\begin{center}
\includegraphics[scale=1.0,clip]{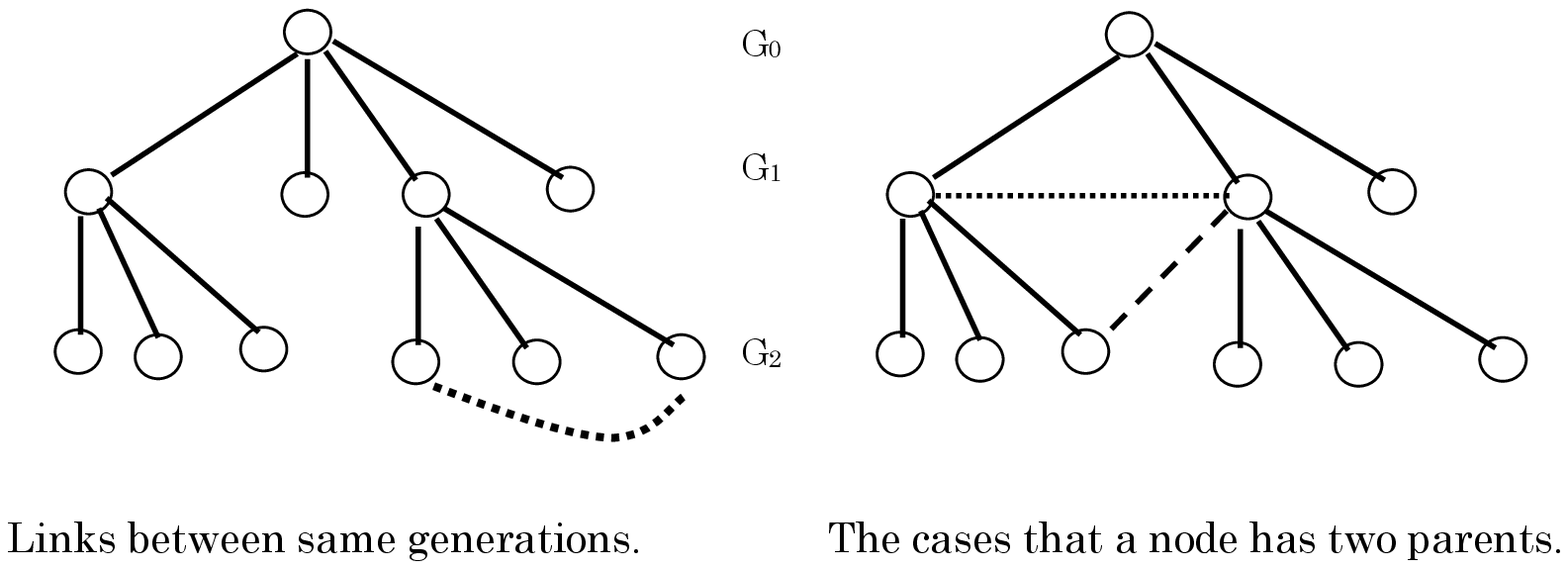}\\
Fig．3.  Two patterns that contribute to  a clustering coefficient in $G_i$ generation. 
The possibility that the right hand pattern occurs is $q$. 　　
\end{center}　

Since each node has $K$ edges from the assumption 4., $C_i$ is generally obtained by 
\begin{equation} C_i= \frac{t_i(q)}{_{K}C_{2}}. \end{equation}
We investigate  two types of contributions to $ t_i(q)$ in order. 
 
First we consider the case in the left hand side of Fig.3, that is to say $ t_i(0) $.  
  When one edge between the same generation $G_i$ is added, 
the average probability that the edge is just one between children nodes with a common parent node in $G_{i-1}$  is 
$_{\overline{k}_{i-1,i}}C_2 / _{n_{i}}C_2 $. 
So as $k_{i,i}$ edges in  $G_i$ are totally added, the number of triangles in $G_i$ become 
\begin{equation}
t_i(0) = \frac{  _{\overline{k}_{i-1,i}}C_2  k_{i,i}n_{i-1}}{ _{n_{i}}C_2 } 
= \frac{ k_{i,i} ( \overline{k}_{i-1,i} -1 )}{  \overline{k}_{i-1,i} n_{i-1} -1 }
=\frac{ k_{i,i} ( \overline{k}_{i-1,i} -1 )}{  \prod_{l-1}^i \overline{k}_{l-1,l} -1 }, 
\end{equation}
since there are $n_{i-1}$ families in $G_i$ and $n_i=n_{i-1} \overline{k}_{i-1,i} $ at $q=0$  in  Eq. (9)  
is used in the last equality. 
We can obtain explicit expressions for first a few $t_i(0)$ ;
\begin{eqnarray}
t_1(0)&=&k_{1,1}, \nonumber \\
t_2(0)&=&k_{2,2} \frac{ \overline{k}_{1,2}-1 }{K\overline{k}_{1,2}-1 }, \nonumber \\
t_3(0)&=&k_{3,3} \frac{ \overline{k}_{2,3}-1 }{K \overline{k}_{1,2} \overline{k}_{2,3}-1 }, \nonumber \\
  \cdots &\cdots&\;\;\;\;\;\;\;\;\;\; \cdots .
\end{eqnarray}

In the other case, the possibility that two parents are linked in their generation $G_{i-1}$ is
$  k_{i-1,i-1} / _{n_{i-1}}C_2 $ 
and the number that  nodes in $G_i$ have two parent nodes is $ q n_{i-1} \overline{k}_{i-1,i}  $. 
Combining these two facts, the contribution $t^{\prime}_i(q)$   to $t_i(q)$ of this case is given by 
\begin{equation} 
t^\prime (q) =(q n_{i-1} \overline{k}_{i-1,i}   ) \frac{k_{i-1,i-1}}{ _{n_{i-1}}C_2}
= \frac{2q\overline{k}_{i-1,i} k_{i-1.i-1}}{n_{i-1}-1}
= \frac{2q\overline{k}_{i-1,i} k_{i-1.i-1}}{ \prod_{l-1}^i \overline{k}_{l-1,l}(i-q)^{1-q} -1},  
\end{equation}
where Eq.(9) is used in the last equation.

Thus  $ t_i(q)$ is finally obtained by 
\begin{equation}
t_i(q)= \frac{1}{_{n_i}C_2} \Bigl( k_{i,i} n_i \times _{\overline{k}_{i-1,i}}C_2 \Bigr)
+  \frac{1}{_{n_{i-1}}C_2} \Bigl( q k_{i-1,i-1} n_{i-1}\overline{k}_{i-1,i} \Bigr),  
\end{equation}
where each term in parentheses shows the contribution from the left hand figure and the right one in the Fig.3, respectively. 

Thus the clustering coefficients $C_i$ in the  generation $G_i$ are obtained by 
\begin{equation}
C_i = \frac{ \overline{k}_{i-i,i} n_{i-1}  }{K(K-1)} 
\biggl( \frac{ (\overline{k}_{i-1,i} -1)(K-1-\overline{k}_{i,i+1})}{ \overline{k}_{i-1,i} n_{i-1} -1 } +
\frac{2q(K-1-  \overline{k}_{i-1,i})}{n_{i-1}-1}
\biggr), 
\end{equation}
where the equation derived from Eq.(4) 
\begin{equation}
k_{i,i}= \frac{n_i}{2} (K-1- \overline{k}_{i,i+1} ) = \frac{n_{i-1}\overline{k}_{i-i,i}  }{2} (K-1- \overline{k}_{i,i+1} ) 
\end{equation}
is used.

By using $n_0=1$ and $\overline{k}_{0,1}=n_1=K$, we can give explicit expressions for first a few $C_i$; 
\begin{eqnarray}
C_1 &=& 1-\frac{\overline{k}_{1,2} }{K-1}, \nonumber \\
C_2 &=&  \frac{\overline{k}_{1,2}}{K-1}  
\Bigl(\frac{ (\overline{k}_{1,2}-1) (K-1- \overline{k}_{2,3}) }{K \overline{k}_{1,2}-1}
+\frac{2q (K-1- \overline{k}_{2,3})}{K-1}     \Bigr),   \nonumber \\
  \cdots &\cdots&\;\;\;\;\;\;\;\;\;\; \cdots .
\end{eqnarray}
Using the homogeneous hypothesis $C_i = \overline{C}=const.$, 
we can express every $ \overline{k}_{i-1,i} $ in terms of the $ \overline{C} $.  
By way of example, we obtain 
\begin{eqnarray}
\overline{k}_{1,2}&=& (K-1)(1-  \overline{C} ),\nonumber \\
\overline{k}_{2,3}&=& (K-1)-  \frac{\overline{C} \bigl( 1-2q(1- \overline{C} \bigr) }{1- \overline{C}}  
\frac{ K(K-1)(1-  \overline{C} ) }{(K-1)(1- \overline{C} )-1}),\nonumber \\
\cdots &\cdots&\;\;\;\;\;\;\;\;\;\; \cdots .
\end{eqnarray}
In general we notice that the following recursion relation is satisfied;
\begin{equation}
A_{i,i+1} = 
\frac{\overline{C}K(K-1)( n_{i-1}-1) 
-2qA_{i-1,i} \overline{k}_{i-1,i} n_{i-1}}{\overline{k}_{i-1,i}(\overline{k}_{i-1,i}-1) n_{i-1}(n_{i-1}-1)}, 
\end{equation}
where
\begin{equation}
A_{i,i+1} \equiv K-1- \overline{k}_{i,i+1}.
\end{equation}
$A_i$ means the number of edges connected between the same generation $G_i$. 
We can numerically solve this recursion relation with respect to $ \overline{k}_{i-1,i}$. 

In order to get numerical value of $ \overline{k}_{i-1,i}$, 
we need to fix three parameters, $ q$, $K $, and $\overline{C} $. 
When  $ \overline{k}_{i-1,i}$ is calculated, $N$ is found from Eq. (9) and (10). 
If  network topology was a tree structure, 
 information would spread over $\sim K^l$ persons at $l$ steps as stated before.  
In the present case, it is inferred that the clustering coefficient has strong influence on the propagation of information. 
Then  the spread of information would be restricted to rather less persons in networks with large $C$.    
We consider how the spread of information is restricted owing to $C$.  
We measure it by propagation ratio $R$ that the ratio of the population $N$ really conveyed in the present case to $K^l$;
\begin{equation}
R=\frac{N}{K^l}. 
\end{equation}
 
We evaluate $R$ at $l=6$ with a wide range of the three parameters $ q$, $K $, and $\overline{C} $ 
within positive  $ \overline{k}_{i-1,i}$.  
They are listed in Table 3. 

\begin{center}
Table 3.  Propagation ratio and $C_{min}$ in parameter space $(q,K)$ at $L=6$ \\
\begin{tabular}{|c|c|c|c|c|c|c| } \hline 
K/q &q=0 &q=0.1 &q=0.2 &q=0.3 &q=0.4 &q=0.5 \\ \hline
$K=10^3$  &0.996-0.229 &0.996-0.032 &0.996-0.020&0.996-0.025 &0.996-0.028 &0.996-0.013\\
          &$C=$0.29 &$C=$0.33 & $C=$0.38  &$C=$0.42 &$C=$0.46 &$C=$0.51   \\ \hline                                                     
$K=10^2$  &0.961-0.035 &0.961-0.037 &0.961-0.025 &0.961-0.029 &0.961-0.016 &0.961-0.018\\
          &$C=$0.28 &$C=$0.32 & $C=$0.37  &$C=$0.41 &$C=$0.46 &$C=$0.50   \\ \hline  
$K=50$  &0.876-0.004 &0.876-0.041 &0.876-0.028 &0.876-0.001 &0.876-0.001 &0.876-0.023\\
         &$C=$0.27  &$C=$0.31 & $C=$0.36  &$C=$0.4 &$C=$0.46 &$C=$0.49   \\ \hline  
$K=28$  &0.82-0.018 &0.82-0.0015 &0.82-0.0054 &0.82-0.0102 &0.82-0.0147 &0.82-0.0175   \\
        &$C=$0.27   &$C=$0.32 & $C=$0.36  &$C=$0.4 &$C=$0.44 &$C=$0.48 \\ \hline  
$K=10$  &0.664-0.046 &0.664-0.041 &0.664-0.039 &0.664-0.006 &0.664-0.006 &0.664-0.030\\
        &$C=$0.21   &$C=$0.25 & $C=$0.29  &$C=$0.33 &$C=$0.39 &$C=$0.42   \\ \hline  
\end{tabular}
\end{center}
 We find the following facts that by observing in entire parameter region of $(q,K)$-space; \\
(1)$0.29<C<0.51$,\\
(2)$N \sim (100 \sim \mbox{a few percent) of } K^{d=6} $,\\
(3)$\overline{k}^j_{i,j}\sim \frac{K}{2}$. \\

Information can spread over a large portion of $K^l$,  even when $C$ has rather a large value.
It is like  "small world network" proposed by Watts and Strogatz\cite{Watt1},\cite{Watt2}. 
From Table 3, we find that a person needs only have about 50 acquaintances 
in order that information can spread over a few hundred million from only a person even at the worst 
with the largest clustering coefficient.  
This condition may be satisfied rather easily in the actual society. 


\section{Some Empirical Networks}
We consider data  on Mixi\cite{Yuta1}  and Bacon Game\cite{Bacon} as suitable networks for our aim. 

\subsection{ Mixi Data}
We can interpret Table 4\cite{Masu1}  to mean  that a person can convey information to how many persons at each generation.  
First three columns in the Table 4 is based on the data described by Masuda\cite{Masu1}. 
The last column is average  propagation coefficients $ \overline{k}_{i-i,i}  $ at 
each generation. 
Every column in the Table 4 except the first one is shown in Fig. 4 where the length means the distance 
between a root, that is, a first person and  a descendant on the graph.   
Notice that the length is also the same as generation number $i$. 
The distribution of the number of new nodes at $G_i$ is a bell-shaped with the peak at the $length=5$. 
Its cumulative distribution has a logistic shape that  shows nearly all of the population are included up to the $length=6$. 
It  drastically rises at the $length = 5\pm1$. 
It is, meanwhile,  speculated that some hubs are brought over to the distribution.     
This suggests that this network is scale free as also suggested by Yuta et al. \cite{Yuta1}.     
The propagation coefficient exhibits a strange behavior at first glance.  
We speculate that this is an incidental event as will be observed in the next subsection. 

\begin{center}
Table 4.  Data on Mixi by ATR group  \\
\begin{tabular}{|c|c|c|c| } \hline 
generation   $i$   & number of new nodes at $G_i$   & number of total nodes & propagation coefficient  \\ \hline 
0 &1 &1 &28 \\ \hline
1 &28 &29 & 9.4642857 \\ \hline
2 &265 &294 & 24.430189\\ \hline
3 &6474 &6768  & 13.792864 \\ \hline
4 &89295 & 96063 & 1.9807044 \\ \hline
5 &176867 & 272930 &0.417692 \\ \hline
6 &73876 & 346806 & 0.167334452\\ \hline
7 &  12362& 359168 & 0.1173758\\ \hline
8 & 1451& 360619 & 0.0978635 \\ \hline
9 & 142& 360761 &0.2112676\\ \hline
10 &30	& 360791 & 0.133333333\\ \hline
11 & 4& 360795 & 1.250 \\ \hline
12 & 5& 360800 & 0.4 \\ \hline
13 &2 & 360802 &  \\ \hline
\end{tabular}
\end{center}

\begin{center}
\includegraphics[scale=1.0,clip]{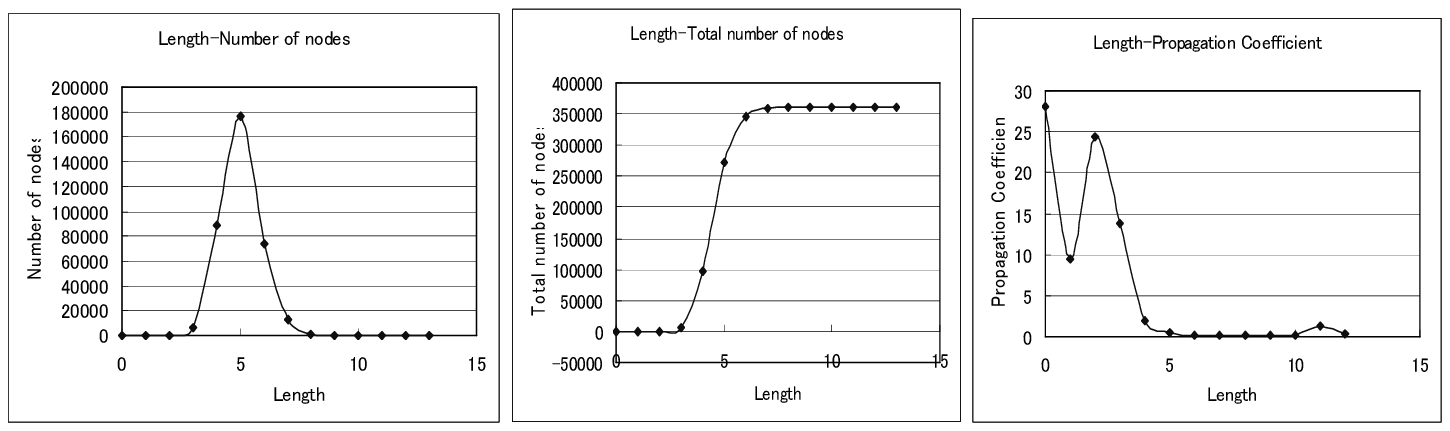}\\
Fig．4  Propagation In Mixi. 　　
\end{center}　

Since we can find $ \overline{k}_{i-i,i}  $  and $n_i$  from Table 4, if $K$ is determined, 
we can evaluate clustering coefficients by Eq (16) for various values of $q$.  
While the average degree is $<K>=10.4$ in the data for Mixi according to Yuta et al.\cite{Yuta1}, 
we assume $<K>=k_{0,1}$ so as to get adequate values for the clustering coefficient.
This is the assumption that the number of edges left from a first one person is $K$. 
Though needless to say that is not always the case, this assumption can derive meaningful clustering coefficients 
in the present case.   
The values of them are given by Table 5. 
This roughly show the following relation holds;
\begin{equation}
C_i \sim 10^{-(G_i+1)}. 
\end{equation}
When $i>3$, the structure of the Mixi network is practically a random graph. 
This also means that $C_i$ is not constant and so the homogeneous hypothesis is rejected in the present case. 
This is due to almost scale free nature of the Mixi network and finite size effects. 
Though considerations in the section 2 is not adequate for such cases, 
they would be adequate to Watts-Strogatz type networks. 
 
\begin{center}
Table 5. Clustering Coefficients for Mixi data based on propagation coefficients with $<K>=k_{0,1}.$ \\
\begin{tabular}{|l|c|c|} \hline 
         &q=0.5  &q=0.8 \\ \hline
$G_1=1$  & none       & none    \\ \hline
$G_1=2$  & $9.6\times 10^{-3\sim4}$ & $1.4\times 10^{-3}$ \\ \hline
$G_1=3$  & $7.6\times 10^{-5}$ & $8.3\times 10^{-5}$ \\ \hline
$G_1=4$  & $7.4\times 10^{-6}$ & $8.8\times 10^{-6}$ \\ \hline
$G_1=5$  & $5.7\times 10^{-7}$ & $7.5\times 10^{-7}$ \\ \hline
$G_1=6$  &    none             & $1.9\times 10^{-8}$ \\ \hline
\end{tabular}
\end{center}

\subsection{ Bacon Game}
We can draw a good deal of information from the web page\cite{Bacon}. 
Fig. 5 displays figures similar to Fig.4 in the case of Bacon game. 
The above figures in Fig.6 are drawn when the starting person is literally Bacon and the below ones  
are done when the starting person is the one with the longest average path length on the actor network. 
They are each other quite alike apart from the position of the peak of the distribution (both ends) 
 or maximum gradient (the middle). 
Key actors would  connect  with all the actors by short distances  so that  
the distance at the peak  is also small. 
Especially the propagation coefficient marks its peak  at the first step for the leading actor Bacon.  
These observations demonstrate that the homogeneous hypothesis is not valid like  the subsection 4.1. 

In these experimental networks, we find that 
$C_i$ does not take a constant value mainly due to finite size effect and almost scale free nature. 
A simple estimation of $C_i$ occasionally leads to some negative value owing to these.  
A problem in which $C_i$ becomes negative when  we take $K=10.4$ in the subsection 4.1 crops up.   
On closer investigation, a finite size effect brings on $\bar{k}_{i-1,i}<1$ and 
hubs alone are connected to extraordinarily many persons. 
These cause negative $C_i$. 

In order to solve the problems, we should consider this subject more minutely 
beyond the homogeneous hypothesis, including distribution of $C_j$ of the individual, 
a degree correlation between neighboring persons and so on.  

\begin{center}
\includegraphics[scale=0.8,clip]{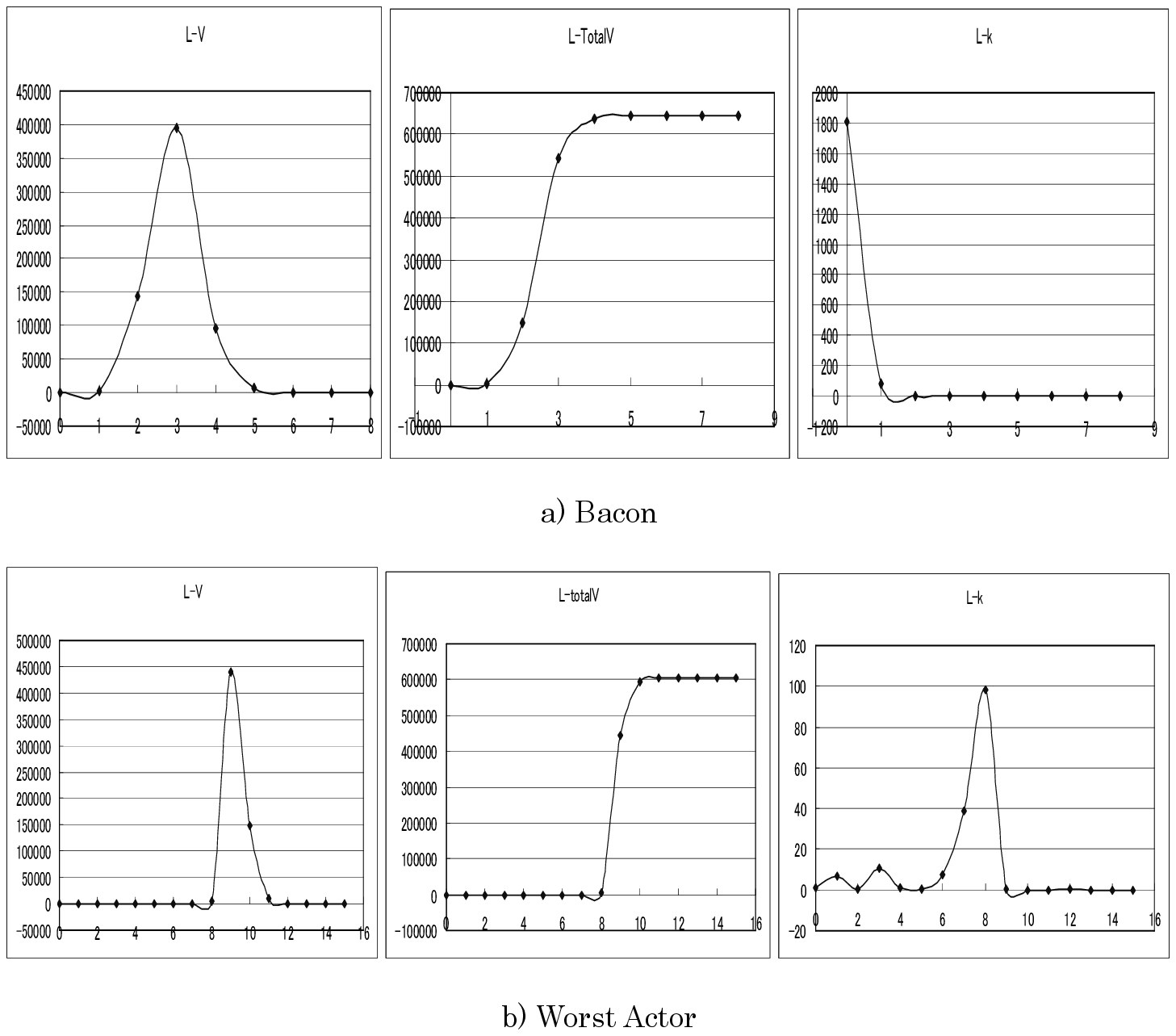}\\
Fig．5  Propagation In Bacon Game. Left: the number of new nodes appeared at a generation $G_i$.  Middle: the number of total nodes appeared up to the generation $G_i$. Right: The propagation coefficient at the generation $G_i$.  　　
\end{center}　

\section{Summary}
In this article we investigate three points about six degrees of separation proposed by Milgram\cite{Milg}. 
First one is the analyses of the Watts-Strogatz type small world network, which is based on analytic expressions for the clustering coefficient and the average path length. 
Second we gave discussions on the  propagation coefficient model based on the homogeneous hypothesis 
for information transmission.
Though empirical networks  do not necessarily support the hypothesis, we turn out  to establish the  formalism to calculate propagation coefficient and so on in the schemes where  information propagates  from generation to generation.  
Third  some experimental networks were investigated where the validity of the homogeneous hypothesis was examined 
so that some  points at issue were made clear.   
Lastly numerical analyses carried out in  these three subjects.  
Knowledge gained through these investigations is summarized as follows. \\

1. The effect of the clustering coefficient on diffusion of information over networks of human relations  
 is not so crucial. 
The effect of clustering coefficient on them only reduces  the population who can receive information in a tree graph 
to a few percentage of it. 
Though there is  a  double-digit decrease in the population who can receive information, each person  has only to 
have dozens  of contacts for six degrees of separation. 
Thus  "six degrees of separation" is not so amazing phenomenon.  \\

2.The homogeneous hypothesis should be made more accurate.   
By way of example, considerations for distribution of the clustering coefficient in every node, 
degree correlation between neighbors as well as degree distribution and so on should be given. 
They are  future issues to be addressed.   \\

3. The finite size effect of networks has to be considered in the analysis of empirical data.\\

They are summaries and future issues of this article. 


\end{document}